# Hydrogen Fluoride Capture by Imidazolium Acetate Ionic Liquid

Vitaly Chaban[1]

1) Instituto de Ciência e Tecnologia, Universidade Federal de São Paulo, 12231-280, São José dos Campos, SP, Brazil

2) Department of Chemistry, University of Southern California, Los Angeles, CA 90089, United States

**Abstract**. Extraction of hydrofluoric acid (HF) from oils is a drastically important problem in petroleum industry, since HF causes quick corrosion of pipe lines and brings severe health problems to humanity. Some ionic liquids (ILs) constitute promising scavenger agents thanks to strong binding to polar compounds and tunability. PM7-MD simulations and hybrid density functional theory are employed here to consider HF capture ability of ILs. Discussing the effects and impacts of the cation and the anion separately and together, I will evaluate performance of imidazolium acetate and outline systematic search guidelines for efficient adsorption and extraction of HF.

---

[1] E-mail: vvchaban@gmail.com

TOC Image

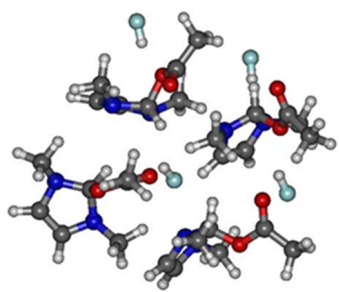 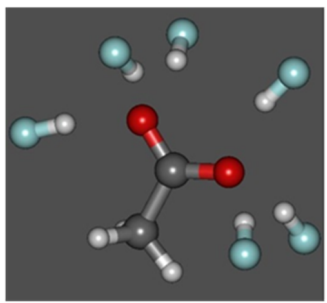

**Research Highlights**

1) Hydrogen fluoride capture ability of imidazolium acetate is researched.

2) Electronic and structure properties are reported.

3) PM7-MD simulations allow describing versatile interactions in these systems.

**Introduction**

Room-temperature ionic liquids (RTILs) are subject to scientific and technological interest due to their unique and versatile properties, such as low melting points and vapor pressures, good thermal and electrochemical stabilities (in many cases), catalytic activities, universal solvation potential with respect to both polar and nonpolar compounds.[1-30] Many RTILs were first considered green. However, currently the environmental friendliness of many of them is being reconsidered with more thorough attention.[1] Properties of RTILs can be tuned for specific applications by changing their ionic components and the structure of each ion. The concept of task-specific RTILs is a global initiative with a technological promise.

Hydrogen bonds play a major role in determination of the RTIL structure, aggregation state, and solvation behavior.[5,15] Hydrogen bonds are usually created between the heterocyclic ring of the cation and polar moiety of the anion. Furthermore, the cation can be functionalized to offer competitive hydrogen bonding sites, e.g. instead of, otherwise inert, hydrocarbon chain.[1] Both cation and anion can participate in solvation and gas capture. The specific behavior depends on the chemical identities of the cation, the anion, and the solute.[3] Excessive generalization must be avoided in this context. Below I enumerate a few important examples where application of RTILs as solvents for extraction was considered.

Pomelli and coworkers[23] showed the potential of RTILs in oil desulfurization finding a similar ordering between experimental solubility and calculated binding energy of small $H_2S$-anion complexes. These complexes did not include the cation, however. Although the correlation has been successfully established, neglecting cation may have dramatic consequence. The effect of the liquid phase can only be captured if both ions are considered in the calculations. It is furthermore better to include more ions than a single ion pair. Damas and coworkers[26] studied the interactions between polar gas molecules ($CO_2$, $SO_2$, and $H_2S$) and ions constituting RTILs at the B3LYP/6-311+G** level of theory. The experimentally observed high $CO_2$ solubility in 1-

butyl-3-methyliidazolium bis(trifluoromethanesulfonyl)imide is caused by weakening cation-anion binding. This weakening increases the free volume in RTIL providing more space to absorb gases (entropic contribution). One can additionally rephrase that $CO_2$ competes with the anion for the cation's binding site. The cation-anion binding decays, whereas the cation-$CO_2$ binding strengthens. Ghobadi and coworkers[31] reported solubility of sulfur dioxide ($SO_2$) and carbon dioxide ($CO_2$) at 1 bar in a few imidazolium-based RTILs. Gas solubility index has been proposed as a ratio of gas-RTIL binding to cation-anion binding normalized per unit of volume. According to this index, the ions must be as bulky as possible for an efficient gas capture.

A comprehensive review paper has been recently contributed by Lei and coworkers[32] summarizing solubility data of $CO_2$, $SO_2$, $CO$, $N_2$, $O_2$, $H_2$, $H_2S$, $N_2O$, $CH_4$, $NH_3$, noble gases and hydrofluorocarbons in RTILs. Solubility data for mixed gases in RTILs is still scarce. Knowing gaseous mixture solubility is interesting in the context of the free volume concept. One gas may foster free volume creation, while another gas may occupy it. Molecular co-solvents are also able to adjust free volume. RTILs exhibit high adsorption potential in relation to many gases and mixtures. An ambitious goal is to replace conventional volatile organic solvents without the need of complicated industrial setups. Jalili and coworkers[28] contributed solubility of $CO_2$, $H_2S$, and their mixture in 1-octyl-3-methylimidazolium bis(trifluoromethyl)sulfonylimide, [$C_8$MIM][TFSI]. The solubility of $H_2S$ was reported to be twice higher than that of $CO_2$ in this RTIL. Electronic structure calculations at the B3LYP/6-311++G(2d,2p) level of theory suggest that binding of $H_2S$ is indeed somewhat stronger, as compared to $CO_2$. Therefore, certain correlation between binding energies and experimental solubility exists. Note that this correlation is relative rather than absolute, i.e. one cannot hypothesize an absolute value of solubility based on the absolute binding energy following a quantum-mechanical description.

The extraction of polar compounds from a nonpolar media constitutes a complicated engineering problem. Hydrofluoric acid (HF) must be continuously removed from oils in petroleum industry. Alkylation is a common process in oil refineries to produce high octane gasoline from isoparaffin olefins.[33] This process employs sulfuric and hydrofluoric acid as catalysts leaving traces of sulfur and fluorine as by-products.[34] Residual concentrations of hydrofluoric acid are enough to cause severe corrosion in pipe lines, storage tanks, containers, valves, etc. HF causes health problems in humans and contaminates environment after the oils are combusted. These drastic problems motivate the search of efficient HF scavenger agents.[33,34] RTILs represent a promising option due to versatility of the exhibited inter-molecular interactions. RTILs can be considered a large and interesting group of solvents for adsorption and extraction of HF from alkylation gasoline.

This Letter starts from binding energy considerations based on the hybrid density functional theory (HDFT)[35] for a variety of very simple gas-phase complexes: one cation, one anion and one HF molecule. The complex exhibiting most favorable interactions will be subject to molecular dynamics investigation at ambient conditions using the recently formulated PM7-MD method.[3,36,37] The PM7-MD method accounts for thermal motion (entropy) effects and, therefore, mimics realistic HF capture by the selected RTIL. Such a combined simulation approach allows to (1) characterize performance of RTILs with respect to HF capture; (2) isolate impacts of the cation and the anion; (3) identify most usable binding sites of RTILs; (4) observe finite-temperature structure of the RTIL-HF system.

**Methodology**

Hybrid density functional theory empowered by the M06 functional,[35] was used to compute non-covalent binding energies. The wave function of the ground electronic state was expanded using the 6-311+G* basis set. As follows from the above basis set notation, polarization and diffuse functions were added to every atom except hydrogen. Polarization and diffuse functions are important to accurately fit an electronic structure of RTILs, in particular that of anion. All electrons were described explicitly. No effective-core potentials were applied.

The M06 HDFT functional accounts for dispersive attraction by introducing an empirical correction.[35] The basis set superposition error was deducted from the interaction energies using the counterpoise method. Three arbitrarily different starting configurations of each complex were used to search for the deepest energy minimum on the potential energy surface using the conjugate gradient geometry optimization algorithm. The reported binding energies correspond to the lowest-energy optimized conformation. The HDFT computations have been conducted using the Gaussian 09 (revision D) quantum-chemistry suite, *www.gaussian.com*.

The PM7-MD simulations[3,36,37] were performed using the five systems (Table 1) following the results of binding energy calculations. All simulations were performed at 300 K. Note that bulk HF exists in the gaseous state at this temperature.

Table 1. Simulated systems, their fundamental properties and selected simulation details. Proper equilibration of all systems was thoroughly controlled by analyzing evolution of many thermodynamic quantities, such as energy components, dipole moments, selected interatomic distances. Note that equilibration of the non-periodic systems occurs significantly faster due to absence of the long-order structure

| # | Composition | # atoms | # electrons | Weight, a.m.u. | Equilibration time, ps | Sampling time, ps |
|---|---|---|---|---|---|---|
| 1 | [Ac+6HF]$^-$ | 19 | 72 | 179.082 | 2.0 | 33 |
| 2 | [IM+6HF]$^+$ | 28 | 86 | 217.177 | 2.0 | 33 |
| 3 | 4[IM][Ac]+2HF | 96 | 264 | 664.749 | 5.5 | 50 |
| 4 | 4[IM][Ac]+4HF | 100 | 280 | 704.761 | 7.0 | 50 |
| 5 | 4[IM][Ac]+6HF | 104 | 296 | 744.774 | 7.0 | 70 |

The PM7-MD method obtains forces acting on every atomic nucleus from the electronic structure computation using the PM7 semiempirical Hamiltonian.[38-41] PM7 is a parameterized Hartree-Fock method, where certain integrals are pre-determined based on the well-known experimental data. Such a solution also allows for effective incorporation of the electron-correlation effects, while preserving a quantum-chemical nature of the method.[41] Therefore, PM7 is able to capture any specific chemical interaction. PM7 is more physically realistic than any existing force field based technique. Note that PM7 features an empirical correction for dispersive attraction.[41] Thus, it can be successfully used to model hydrocarbon chains, which are abundant in ionic liquids. The accuracy and robustness of PM7 as applied to thousands of versatile chemical systems was demonstrated by Stewart elsewhere.[38-41]

The derived forces are coupled with the initial positions of atoms and randomly generated velocities (Maxwell-Boltzmann distribution). Subsequently, Newtonian equations-of-motions can be constructed and numerically integrated employing one of the available algorithms. This work relies on the velocity Verlet integration algorithm. This integrator provides a decent numerical stability, time- reversibility, and preservation of the symplectic form on phase space. Due to rounding errors and other numerical inaccuracies, total energy of the system is not perfectly conserved, as in any other MD simulation method. Temperature may need to be adjusted periodically by rescaling atomic velocity aiming to obtain the required value of kinetic energy with respect to the number of degrees of freedom. This work employs a weak temperature coupling scheme[42] with a relaxation time of 25 fs, whereas the integration time-step equals to 0.25 fs. The integration time-step is relatively small, since the HF molecules move very quickly at 300 K. Larger time-step would result in poor total energy conservation.

More details of the present PM7-MD implementation are described elsewhere.[3,37] The method has been successfully applied to address problems of ionic liquids[3,37] and nanoparticles.[36] Cation and anion effects were quantified using point charges, bond orders and energies. Local structure of the liquid-matter systems was characterized using a set of radial distribution functions (RDFs). The RDFs were calculated using simple in-home tools along the sampling stage of each PM7-MD trajectory (Table 1).

Prior to quantum-chemical analysis, geometries of systems I and II were optimized. Annealing PM7-MD simulations[36] were performed on each system including heating up to 300 K during 2.0 ps, equilibrium simulation at 300 K during 2.0 ps and cooling down to 0 K during 2 ps. After the described procedure, the geometries of the systems were optimized using the eigenfollowing algorithm.

**Results and Discussion**

Ab initio derived binding energies and hydrogen bond lengths are summarized in Table 2. These data offer an initial assessment of the HF capturing performance by ionic liquids. Driven by the anticipation that the anion plays more important role in the HF capture than the cation, anions of various nature are scanned, whereas the cations are based on the imidazole ring, except one featuring pyridine ring. The largest binding energy (per mole of ion pairs), 88.8 kJ mol$^{-1}$, is exhibited by 1-vinyl-3-butylimidazolium trifluoroacetate. It is unlikely that fluorine atoms are particularly useful for the HF capture, since 1-butyl-3-methylimidazolium acetate, [C$_4$C$_1$IM][Ac], features a similar binding energy, 80.0 kJ mol$^{-1}$. The observed difference comes largely from the vinyl side chain (modest but systematic interaction energy increase due to van der Waals attraction). 1-butyl-3-methylimidazolium benzoate appears also a little more successful, 84.0 kJ mol$^{-1}$. Note that the provided binding energies correspond to one mole of ion pairs rather than to a unit of volume. Bulkier ions decrease specific density of RTILs. Therefore, if the binding energies are converted per unit volume (employing density of bulk RTIL), the numbers will likely change in favor of 1-butyl-3-methylimidazolium acetate. RTILs containing tetrafluoroborate, hexafluorophosphate and even bis(trifluoromethanesulfonyl)imide are significantly less attractive for hydrogen fluoride. Failure of the bis(trifluoromethanesulfonyl)imide anion must be remarkable, since it performs well for CO$_2$ binding in conjunction with the imidazolium-based cations.

Table 2. Binding energies and closest-approach distances for complexes composed of an ion pair (RTIL) and HF molecule in vacuum. All binding energies are already corrected for basis set

superposition error. The computations have been conducted using the M06 HDFT method, as introduced in the methodology section

| Cation | Anion | $E_{RTIL-HF}$, kJ mol$^{-1}$ | Cation-HF distance, Å | Anion-HF distance, Å |
|---|---|---|---|---|
| 1-butyl-3-methylimidazolium | BF$_4$ | 53.1 | 2.89 | 1.62 |
| 4-methyl-N-butylpyridinium | BF$_4$ | 77.0 | 2.98 | 1.57 |
| 1-benzyl-3-methylimidazolium | PF$_6$ | 54.3 | 2.37 | 1.62 |
| 1-butyl-3-methylimidazolium | CH$_3$COO | 80.0 | 2.13 | 1.50 |
| 1-vinyl-3-butylimidazolium | CF$_3$COO | 88.9 | 2.26 | 1.46 |
| 1-benzyl-3-methylimidazolium | (CF$_3$SO$_2$)$_2$N | 40.0 | 2.95 | 1.69 |
| 1-Butyl-3-methylimidazolium | C$_6$H$_5$COO | 84.0 | 2.69 | 1.42 |

The closest-approach distances for HF are systematically smaller in the case of anions. This happens because HF is oriented to the anion by its hydrogen atom (Figure 1). That is, a small atom is able to approach closer. In turn, the cations are coordinated by fluorine atoms, which are larger. Closer approach means stronger interaction. HF readily engenders hydrogen bonds with all anions, but not with the cations. The only probable exception is [C$_4$C$_1$IM][Ac], where the distance equals to 2.13 Å. The lack of hydrogen bonding of HF with the cations may occur due to the presence of hydrogen bonds with the anion, i.e. the mutual orientation of ions and molecules does not allow HF to maintain two hydrogen bonds per one molecule.

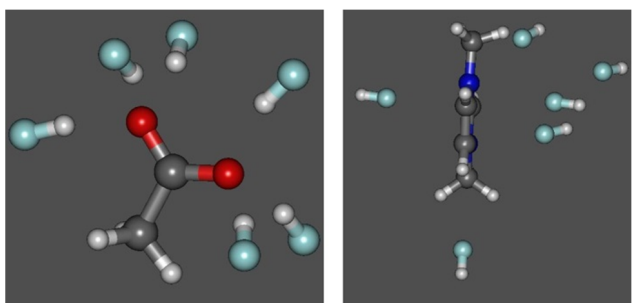

Figure 1. Thoroughly optimized (see methodology) geometries of the two charged ion-molecular complexes. Left: acetate anion surrounded by six hydrogen fluoride molecules (system I). Right: 1,3-dimethylimidazolium cation surrounded by six hydrogen fluoride molecules (system II). Oxygen atoms are red, carbon atoms are grey, hydrogen atoms are white, fluorine atoms are sky blue, nitrogen atoms are blue. The optimized complexes clearly suggest that the anion is a better attractor for hydrogen fluoride.

The careful analysis of binding energies and inter-molecular distances indicates that [$C_4C_1$IM][Ac] is the most interesting candidate for further examination. Bulkier ions exhibiting just insignificantly higher binding energy look less attractive, since the amount of those ions will be smaller than the amount of the [$C_4C_1$IM][Ac] ions in the same volume of liquid. Simulation of the butyl hydrocarbon chain is possible in PM7-MD, but is not sufficiently interesting, since the fatty chain does not directly participate in the HF binding. The simplification can, therefore, be introduced by substituting the butyl chain by the methyl chain. In the following, I report and discuss PM7-MD simulations of the HF capture in [$C_1C_1$IM][Ac].

Molecular configurations depicted in Figure 1 suggest that the [Ac] anion is much better coordinated by the HF, as compared to the [$C_1C_1$IM] cation. This observation is in concordance with the M06 HDFT consideration (Table 2). It is also confirmed by energy of formation analysis (Figure 2). Interaction of [Ac] with the six HF molecules is more than twice stronger despite the fact that [Ac] is smaller than [$C_1C_1$IM]. Interestingly, the geometry of the negatively charged complex relaxes faster, 400 vs. 700 eigenfollowing iterations. Strong interactions normally favor quick geometry optimization.

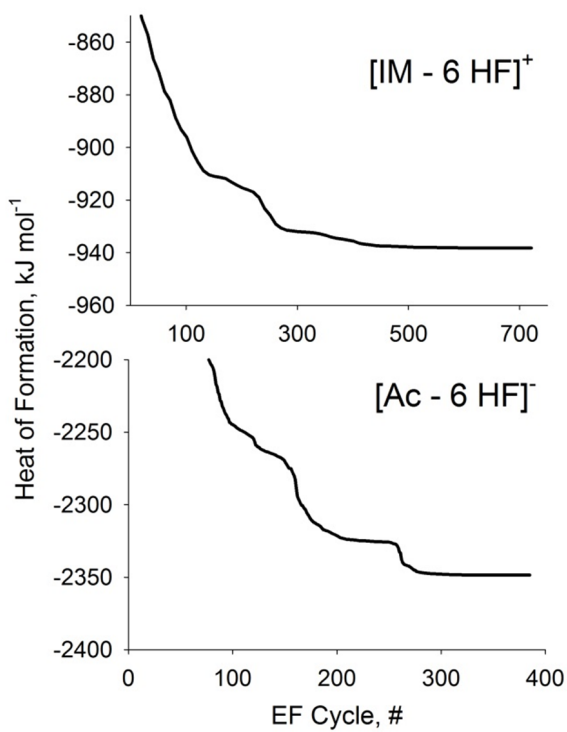

Figure 2. Energy of formation evolution upon eigenfollowing geometry optimizations of systems I-II. The absolute value of formation energy in the negatively charged system remarkably exceeds that in the positively charged system. The energies are provided per mole of the considered systems, rather than per mole of atoms or electrons.

The effect of the cation and the anion on the HF molecules can be conveniently rationalized by the anisotropy of charge density. Figure 3 provides electrostatic potential (ESP) point charges on fluorine and hydrogen atoms in the optimized [Ac+6HF]$^-$ and [IM+6HF]$^+$ complexes. The ESP charges on the fluorine atoms are almost steady, whereas the ESP charges on the hydrogen atoms fluctuate to a larger extent. The particular charge is determined by whether the HF molecule occupies the binding site of the ion or is linked to another HF molecule. The H-F bond is more polar in the [IM+6HF]$^+$ complex. It occurs thanks to a somewhat larger electron density shift from the [Ac] anion. Indeed, an average charge on the HF molecule in [Ac+6HF]$^-$ equals to -0.14e. One can rephrase that HF effectively acts as a very weak anion. Compare, an average charge on the HF molecule in [IM+6HF]$^+$ equals to -0.04e. Thus, in both cases HF pulls some electron density onto itself, due to highly electronegative fluorine, but the effect is at least thrice more pronounced in the negatively charged system. Coupling HDFT and PM7-MD together, one concludes that the anion plays definitely more important role in the HD capture, as compared to the imidazolium-based cation. It is not likely that additional functionalization of the imidazolium-based cation can change the observed behavior. Anions must be tuned and screened to pick up the most successful RTIL.

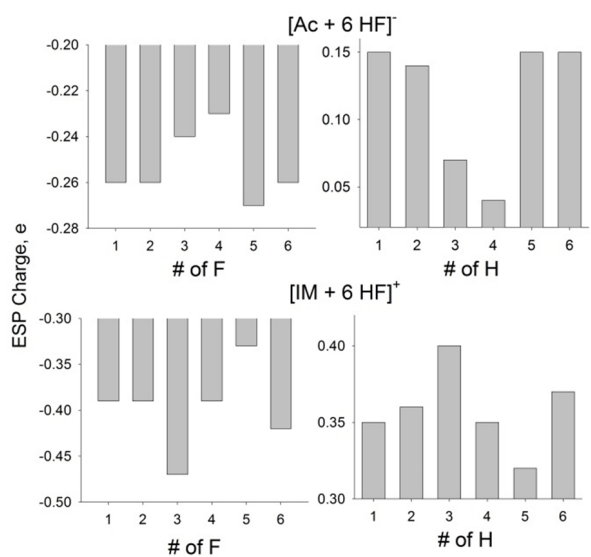

Figure 3. Point charges residing on the hydrogen fluoride molecules obtained via electrostatic potential (ESP) fitting in the positively and negatively charged complexes. Numbers along the x-axis refer to the number of the HF molecule in the simulated system.

Larger simulation setups (systems III-V, Figure 4) are required to consider many-body effects. This investigation considers two, four, and six HF molecules per four $[C_1C_1IM][Ac]$ ion pairs. All these systems are electrostatically neutral.

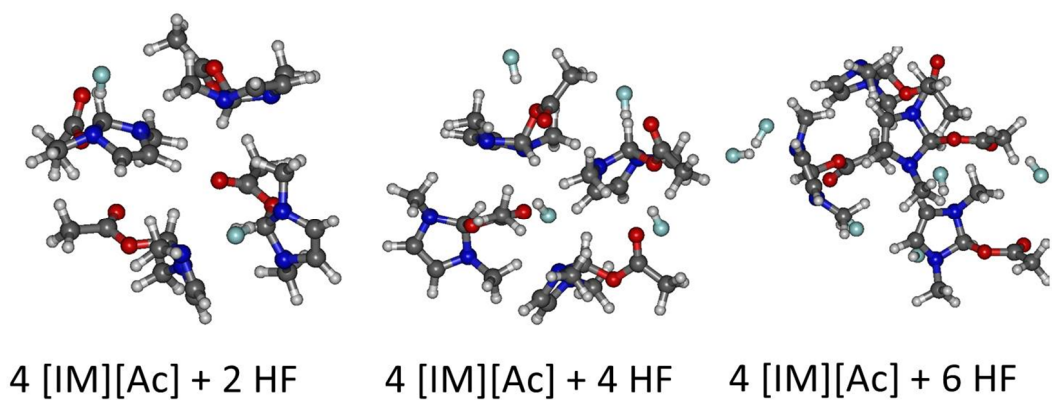

4 [IM][Ac] + 2 HF    4 [IM][Ac] + 4 HF    4 [IM][Ac] + 6 HF

Figure 4. Immediate ion-molecular configurations after molecular dynamics equilibration of systems III-V. Oxygen atoms are red, carbon atoms are grey, hydrogen atoms are white, fluorine atoms are sky blue, nitrogen atoms are blue. The location of the HF molecules indicates their preferential binding sites, which will be systematically described using radial distribution functions below.

Figures 5-6 depict RDFs for the most important atom-atom interactions. The sharp peaks at 2.0-2.1 Å indicate hydrogen bonding engendered by hydrogen of HF and oxygen of carboxyl group. Interestingly, this short-distance peak is absent in system III, the one with the lowest content of HF. The first maximum in system III is observed only at 4.4 Å. Unlike system I, competition for the carboxyl group exists between the imidazolium-based cation and the HF molecules. At lower concentration of HF, [$C_1C_1IM$] appears more successful (Figure 4). However, HF also forms hydrogen bonds at its higher concentrations. Figure 6 univocally confirms the supposition regarding strong cation-anion interactions at finite temperature. Binding of HF to [$C_1C_1IM$] is rather weak with all significant peaks at and beyond 3 Å. Such a spatial separation does not imply a strong interaction.

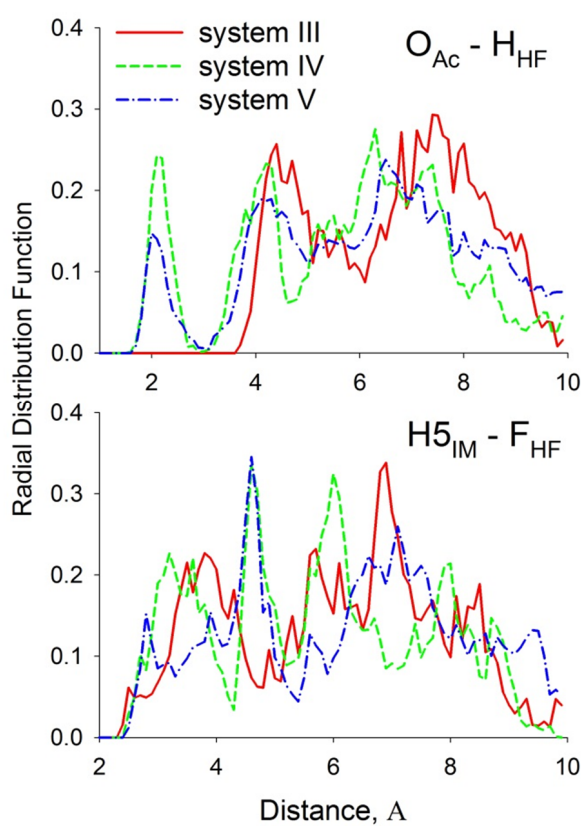

Figure 5. Radial distribution functions derived for (top) oxygen of carboxyl group and hydrogen of HF; (bottom) intrinsically acidic hydrogen of the imidazole ring, H5, and fluorine atom of HF. The selected interaction sites constitute primary targets upon HF capture.

The major difference between the HDFT calculations and the PM7-MD simulations is neglect of (1) entropy effects and (2) many-body effects in the former. HF is a very mobile molecule at 300 K which is difficult to capture. Even if HDFT suggests a strong binding both with the anion and the cation, finite-temperature RDFs indicate that many HF molecules are outside the first solvation shell of the acetate anion. To certain extent, the cation can be viewed as an obstacle, which harms performance of the [Ac] anion. Note, however, that namely [$C_1C_1$IM] is responsible for a liquid aggregation state of [$C_1C_1$IM][Ac] at ambient conditions. An interesting option is to consider fully ionized dicarboxylic acid anions, such as that of oxalic acid, $^-$OOC-COO$^-$. This anion possesses two equivalent binding sites, which are attractive to HF. Due to a bulky shape of the imidazolium cations, they will unlikely be able to occupy both of these sites, whereas small HF molecules can penetrate there much easier.

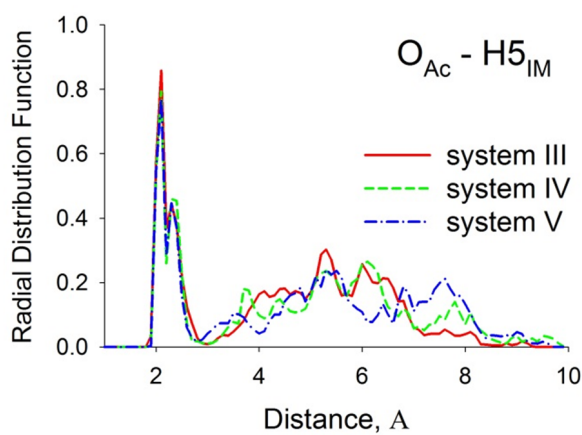

Figure 6. Radial distribution functions derived for oxygen of carboxyl group and H5 hydrogen of imidazole ring. A strong hydrogen bond is observed between the selected sites with a length of 2.1 Å. This hydrogen bond is depicted as a solid line in Figure 4.

**Conclusions**

Hybrid density functional theory and PM7-MD simulations at 300 K were employed to probe a range of RTILs as prospective scavengers in relation to hydrogen fluoride. Based on the binding energy analysis and practical considerations, 1,3-dimethylimidazolium acetate was picked up for a more detailed investigation. The anion, acetate, is of key importance for HF capture. Imidazolium-based cation is fairly useless, but it helps to maintain acetate in the liquid state. Usage of an inorganic cation would have resulted in a solid salt, which can adsorb HF only by means of its tiny surface. Therefore, a cation in combination with acetate is not important as long as it achieves a low melting point.

The reported results and guidelines inspire knowledgeable search of scavengers for hydrogen fluoride, as well as other aggressive and poisonous agents.

**Acknowledgments**